\documentclass[12pt]{article}
\usepackage{epsfig}
\usepackage{cite}
\usepackage{./mcite}
\usepackage{url}

\usepackage{array,tabularx,epsfig,mathrsfs,graphicx,rotating}
\usepackage{ifthen}
\usepackage{fancybox}
\usepackage{amsfonts}

\newcommand{\beq}{\begin{equation}}
\newcommand{\eeq}{\end{equation}}

\newcommand{\bg}{\rho^{\mathrm{bkg} } }
\newcommand{\ks}{K_S^0}
\newcommand{\tp}{\Theta^{+}}

\chardef\til=126
\newcommand{\mev}{{\,\mathrm{MeV}}}
\newcommand{\gev}{{\,\mathrm{GeV}}}
\chardef\usc=95

\begin{document}

\clearpage
\pagestyle{empty}
\setcounter{footnote}{0}\setcounter{page}{0}%
\thispagestyle{empty}\pagestyle{plain}\pagenumbering{arabic}%

\hfill  ANL-HEP-PR-07-47 

\hfill  July 9, 2007


\vspace{2.0cm}

\begin{center}

{\Large\bf
On influence of experimental resolution on \\ the statistical significance of a signal: \\
implication for pentaquark searches
\\[-1cm] }

\vspace{2.5cm}

{\large S.V.~Chekanov~$^a$
and B.B.~Levchenko~$^b$

\begin{itemize}
\itemsep=-1mm

\normalsize
\item[$^a$]
\small
HEP Division, Argonne National Laboratory,
9700 S.Cass Avenue, \\
Argonne, IL 60439
USA

\normalsize
\item[$^b$]
\small
Skobeltsyn Institute of Nuclear Physics, Moscow State University, \\
119992 Moscow, Russian Federation

\end{itemize}
}

\normalsize
\vspace{1.0cm}


\vspace{0.5cm}
\begin{abstract}
An analytical relationship between the statistical significance
of an observed signal and the signal width  in case
of a large 
background was  obtained.  It can help to explain
why high-energy experiments  may have different conclusions on the existence of
new  particles. We illustrate our approach
using the experimental data on searches for the $\tp (1530)$ pentaquark state.
The obtained relationship is also useful 
for planning of future  experiments designed to search for signals of new particles
in invariant-mass distributions.
\end{abstract}

\end{center}

\newpage
\setcounter{page}{1}


\section{Introduction}

With increase of the energy of colliding beams, high-energy experiments
have to deal with the problem of a large combinatorial
background which makes searches for new 
narrow states\footnote{We define narrow states as the states which have the 
natural width less than $10^{-2}\cdot m$, where $m$ is a particle mass.}
in invariant-mass
distributions to be rather difficult.
This was the case with the top-quark  discovery, searches
for new charmonium states or pentaquark searches.
Usually, observed signals do not have high  statistical significance,
and the decision has to be made about whether the observed peak
has a sufficient significance to claim the observation, or
it should be disregarded.

The situation with particle searches could be rather complicated and confusing,
especially when several similar experiments have different conclusions
on a  signal existence. In such cases, instrumental differences
between these  experiments are important  to know.
For the signal reconstruction in invariant-mass distributions, 
it becomes crucial to understand the influence of experimental
resolution on the statistical significance of a signal.
In this paper, we have obtained a relationship between the statistical significance
of a signal and its width. For narrow states, the peak width
is mainly determined by the experimental resolution which can be different for different experiments.
Therefore, the obtained relationship could clarify some  cases when several experiments
observe a signal, while others do not see it.

To be more specific, let us remind that two similar collider
experiments at DESY (Hamburg)~\cite{Chekanov:2004kn,Aktas:2006ic}
and two similar fixed-target experiments at
IHEP (Protvino)~\cite{Aleev:2004sa,Antipov:2004jz} do not agree on the existence of 
an exotic state with the mass close to
1530 MeV, which may be interpreted as a 
strange pentaquark (PQ) state $\Theta^+(1530)$~\cite{Diakonov:1997mm}.
While the ZEUS~\cite{Chekanov:2004kn} and SVD-2~\cite{Aleev:2004sa}
experiments claim the observation of a narrow
peak near 1530 MeV in the decay mode $\tp  \to p\ks$,
the H1 and SPHINX collaborations do no see it.

It should be mentioned that the $p\ks$ decay channel is not exotic,
since the standard $\Sigma$ baryons can decay to this channel as well. 
However, the observed peak position and the narrow width  are
very close to the theoretical expectation for the $\tp$ 
state~\cite{Diakonov:1997mm}.
In addition, a $\Sigma$ state near 1530 MeV is unknown~\cite{Yao:2006px} 
and it was never seen
in previous exclusive reactions where experimental conditions were 
favorable for the observation of the $\Sigma$ states.
There are several indications  that the observed particle has different
properties from those of the usual baryons produced in 
quark and gluon fragmentation~\cite{pqpro,Azimov:2007hs,*Chekanov:2005wf,*Chekanov:2005uz}. 

In this paper, we illustrate 
the influence of the track resolution on 
the statistical significance using the $\tp \to p\ks$ decays.
Our  analysis is
rather general and can be used to establish a relationship between the width
of the observed  signal and its statistical significance in any experiment where a
large background is expected.

\section{Influence of inactive material on  the $p\ks$ \\ reconstruction}
\label{width}

First, let us study the effect of an  inactive material in front of a tracking
device on the reconstructed mass resolution for the $\tp \to p\ks$
decay channel. Clearly, the reconstructed width depends on 
the quality of a tracking device itself, as well as  on the software reconstruction.
For simplicity of our toy simulation, we ignore these facts completely;
our set-up represents an ideal case when the mass resolution is
mainly due to unrecoverable loss of the information
on the original particle momenta after rescattering on 
inactive material. For example, a beam pipe and/or
the inner wall of a tracking device represent an inactive material
leading to a degradation of the track resolution.
In case if a micro-vertex detector is not used as an active
detector in the reconstruction of
the $\ks$ and the proton,
the amount of material further increases and,  
as we show below, the penalty is a significant worsening of the
resolution for the $\tp$ reconstruction. 

For this study, a {\sc Geant}~\cite{Agostinelli:2002hh} 
simulation was used. We made  a simple
detector set-up which consists of an aluminum target followed by a tracking chamber.
Each simulated event  consists of a single initial $\tp$ particle 
with the momentum vector perpendicular to the target, which is
located 1~cm away from the injection point.  
Only the $p\ks$ decay mode was considered.
The initial momentum of $\tp$ was $1\gev$. In this case,
the proton absolute momentum is 
below $1\gev$,  ensuring that the $dE/dx$ identification
is possible. We assume no energy loss in the tracking simulation.

Below we will study the dependence of the  reconstructed width, $w$,  on the amount of material. 
We use the word ``width''  to denote the standard deviation of a Gaussian distribution
used to describe the mass resolution.

The  $\ks$ mesons from the $\tp$ decays interact with
the aluminum target and then decay into $\pi^+\pi^-$. 
Our algorithm calculates the invariant mass of two
oppositely charged tracks  to reconstruct the $\ks$.
For the protons, any track
which is not associated with the pions from the $\ks$ is used.
Then, $\ks$ is
combined with a proton track to reconstruct the $\tp$ mass.
Table~\ref{tab1} shows  the width $w$ for $\ks$ and $\tp$,
when reconstructed tracks are used (the second and third columns), 
or when the true information on the pion  momenta is used (the last column).
The latter reconstruction  is used to illustrate the effect of the proton reconstruction to the width of $\tp$.
Also shown is the width of  the $\tp$ state  when 
the $\ks$ mass is fixed to the PDG value~\cite{Yao:2006px}.
The latter approach was used in most experiments in order to improve
the mass resolution.

\begin{figure}
\begin{center}
\includegraphics[height=14cm]{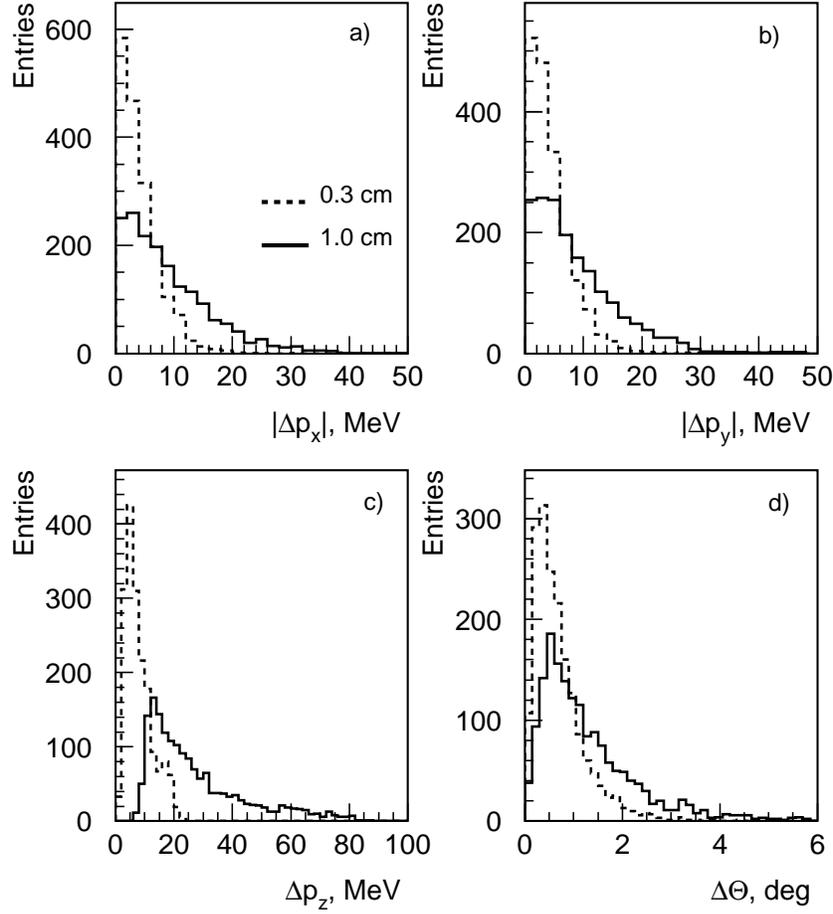}
\caption
{
The distributions of  $x$, $y$ and $z$ components of the proton 
momentum difference between the true and the reconstructed proton momentum,
$\Delta p_i = |\vec{p}_{tr}|-|\vec{p}_{r}|$, after interactions
of the protons from the $\tp$ decay on an aluminum 
target of thickness $L=$0.3~cm
and 1.0~cm. The distributions were obtained using a {\sc Geant} simulation. 
Here  $\vec{p}_{tr}$ is the true momentum of the proton  
originated from a $\tp$ decay and  
$\vec{p}_{r}$ is the   
proton momentum reconstructed after the target;  d) 
distributions of the polar angle
between the momenta $\vec{p}_{tr}$ and  $\vec{p}_{r}$.
}
\label{fig0}
\end{center}
\end{figure}

An additional material in front of a tracking device leads  
to a significant rescattering, as illustrated in Fig.~\ref{fig0}. Obviously,
this is reflected in worsening of the $p\ks$ resolution. In addition, this may affect
the proton-track reconstruction efficiency, 
since a reconstruction program has to make the decision about
whether the proton track belongs to a ``primary-track'' category or not
(assuming that the  $\tp\to p\ks$ is a strong decay).  
However, without a realistic Monte Carlo simulation, to estimate such 
an effect is rather difficult.

Even although our simulation represents a rather simplified situation,
ignoring many details of a realistic track reconstruction,  
the resulting width values of $\ks$ and $\tp$ are rather sensible.
We illustrate this using two similar experiments at HERA, ZEUS and H1.
The ZEUS $p\ks$ resolution (from a Monte Carlo simulation)
is $2.5\pm 0.5\mev$, while the amount of material
in front of the central tracking is $0.3$ cm~\cite{deut} of aluminum. 
This agrees with $2.93\mev$ resolution  given in 
the Table~\ref{tab1} (forth column for $L=0.3$ cm).

For the $\ks$,
ZEUS fits the $\pi^+\pi^-$ peak using two Gaussian
distributions with a common peak position.
The first Gaussian has the width of about $3.5\mev$ 
and describes the main peak, while the second Gaussian with the
width of $7\mev$  describes the tails
(which typically contain about 27\% of all 
reconstructed $\ks$ candidates) \cite{Chekanov:2004kn}.
Thus, the total mass resolution is $4-5\mev$ and it can reasonably be 
approximated by the first Gaussian.
As a cross check, the Half Width at Half Maximum 
(HWHM) was calculated using the plot shown in \cite{Chekanov:2004kn} 
to restore the Gaussian width of the signal. The width was found to be 
around $4\mev$, and agrees with the single-Gaussian width for $\ks$ given in a 
previous ZEUS publication~\cite{Chekanov:2003wc}. 
The ZEUS $\ks$ mass resolution reasonably agrees with the $4.23\mev$ width given in
Table~\ref{tab1} for $L=0.3$~cm.

In case of H1, the $\ks$ mass peak has the width of $8\mev$, 
two times larger than for  ZEUS.
This was obtained by analysing the HWHM of the $\ks$ mass distribution
shown in \cite{Aktas:2006ic}. The quoted RMS for the $\pi^+\pi^-$ mass spectrum
is $9.2\mev$ \cite{Adloff:1997ym}. 
This width is equivalent to twice larger amount of inactive material,
as it follows from Table~\ref{tab1}. Indeed, H1 has an additional detector
in front of the central tracking\cite{Aktas:2006ic} 
which was absent in ZEUS case.

Table~\ref{tab2} demonstrates the relationship between  the mass resolutions
and the conclusions from the $\tp$ searches in the experiments
which have studied identical reactions.  It is obvious that the experiments
with positive evidence for the $\tp$ state have better tracking 
resolution\footnote{ZEUS has also higher statistics for the $p\ks$ combinations compared to H1.} than those which have null evidence. 
The $\ks$ resolution can be translated to the resolution
expected for the $\tp$ state using Table~\ref{tab1}.  
For example, for the quoted above values,
the resolutions of $\tp$ are  $2.9\mev$ and 5--6~MeV for ZEUS and H1 case, 
respectively. These numbers from our simulation agree with those quoted 
in ZEUS and H1 papers~\cite{Chekanov:2004kn,Aktas:2006ic}.

It is conceivable that the different conclusions on the existence
of the $\tp$ state may be related to differences in 
the mass resolution, which
determines the observed peak width in case of narrow resonances.
Below we will study the dependence of the statistical
significance of the signal observation on its width 
in case of a large background. 

\section{Statistical significance and the peak width}
\label{stat}

There are several ways to define the statistical significance, $S$,
of an observed signal. We will use the most practical definition,
which is often used in experimental papers. We define $S$ through the 
ratio of the total number of reconstructed signal entries, $N$, divided
by its error $\delta N$, i.e. as $S\,=\,N/\delta N$. The numbers $N$ and 
$\delta N$ can be found from a fit using a Gaussian plus a background
function. 
The standard deviation, $\sigma$, provides a best estimation for 
the statistical error $\delta N$ \cite{blobel:1998:statistik}
and therefore often the statistical significance is
quoted in the form $S\cdot\sigma$.
We do not use the
definition of $S$ in terms of the probability of the  background  to
fluctuate to a ``signal''  with a certain number of observed events,
since such a definition would require a significant computational time for
the studies to be discussed below.

The statistical significance of a signal observation depends on several factors, and the most
crucial are:  1) the production cross section; 2) the background level under the peak;
3) experimental resolution; 4) the shape of the background.
Below we  will consider  somewhat simplified situation,  
assuming that a Gaussian-like  signal
is located  on a smoothly falling convex-like background, 
since this is the most common situation for many experiments.
The location of the signal is assumed to be known. 
The case when the signal is
expected on a background hump is more difficult and, in some cases,
could  be avoided  by changing selection cuts.

The statistical significance $S$ is usually expressed via
the numbers  of the signal events and the background events. Thus,  
such a definition does not contain explicitly the signal width.
Therefore, we will use different variables for our analysis.  
We  define $\bg$ as the density of the 
background at the mass region where the signal is expected.
It is calculated as the number of combinations at the expected
location of the peak position divided by the bin width.
The  $\bg$  does not
depend very strongly on the shape of the background in the signal region,
assuming that the  signal is narrow and the background
is sufficiently smooth. This variable is determined by the available
statistics and combinatorial background contributing 
to the invariant mass.

Another independent variable, $f$,
is the ratio  of the number of {\em expected}  signal events $N_s$  
over the background density,  $f=N_s /\bg$. 
This variable can be calculated from the expected signal cross section, 
available luminosity and acceptance.
It should be noted that we do not use the definition of the fraction
as the number of signal events divided by the number of background
events under the peak. In this case, such a definition will have
a dependence on the peak width, $w$. This has to be avoided;
the peak width will be another independent variable. 

Our task is to obtain a relationship between $S$ and three
independent variables, $\bg$, $f$ and $w$. 
For this, we have carried out 
several  Monte Carlo  ``experiments'' by generating
background distributions combined with a Gaussian signal with a fixed peak
position at $1520\mev$.
The shape of the background was taken from~\cite{Chekanov:2004kn,Aktas:2006ic}; in fact,
such a threshold shape is rather typical for many searches.
The width of the Gaussian distribution was varied between  3 MeV and 14 MeV.
The fraction of the signal events, $f$,
and the background density $\bg$ were also varied in a wide range. 
For each generated distribution with the background plus the Gaussian signal, 
the  statistical significance was
estimated by fitting the peak using exactly the same functions as those
used to generate the mass spectra. To reduce the number of unstable fits,
the peak position was fixed to the expected value $1520\mev$ during the fit procedure.

\begin{figure}
\hspace*{-15mm}
\begin{minipage}[h]{.5\textwidth}
\includegraphics[height=6.7cm,width=8.5cm]{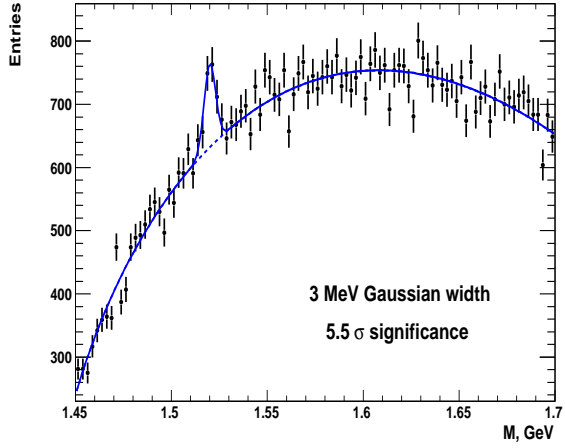}
\includegraphics[height=6.7cm,width=8.5cm]{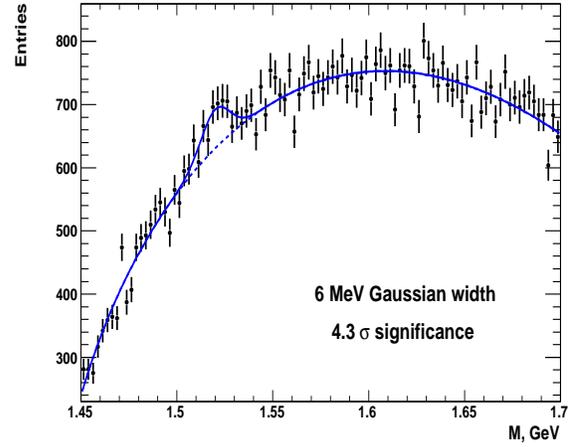}
\end{minipage}  \hspace*{+7mm}
\vspace*{-2mm}
\begin{minipage}[t]{.5\textwidth}
\includegraphics[height=6.7cm,width=8.5cm]{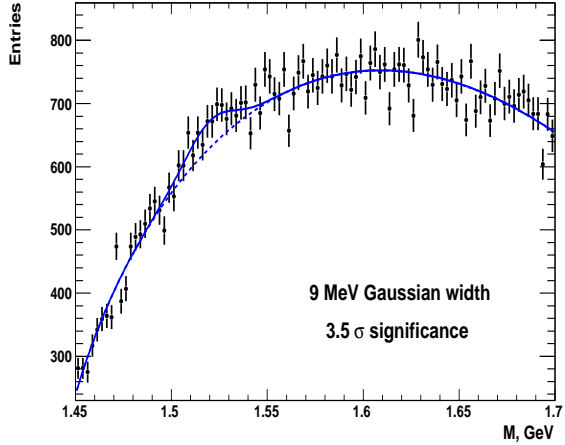}
\end{minipage}
\begin{center}
\caption
{
The generated mass spectra using a threshold function plus a Gaussian
signal at $1520\mev$. The total number of simulated
background events is 65000, $\bg=617/0.0025=246800\gev^{-1}$ and $f=0.0017\gev$.
The widths of the generated signals were set as indicated in the figure.
Also shown are the  statistical significances  $S$ of the extracted signals 
using a fit with a Gaussian
plus a threshold background.
}
\label{fig1}
\end{center}
\end{figure}

Figure~\ref{fig1} shows the  mass distributions simulated using a threshold
function plus a Gaussian with the peak widths 3, 6  and 9 MeV, respectively.
The statistical significance, under the assumption that the location of the
signal is known, was estimated by performing a Gaussian fit plus the threshold
function. The fraction of the signal events, as well as the total number
of the background events, was the same  in all cases.
It is seen that the statistical significance decreases with increase of the
peak width; for $w=3\mev$, one can claim a discovery, while
for $w=9\mev$, the statistical significance of the observation is  low.

\begin{figure}
\begin{center}
\begin{minipage}[c]{0.48\textwidth}
\mbox{\epsfig{file=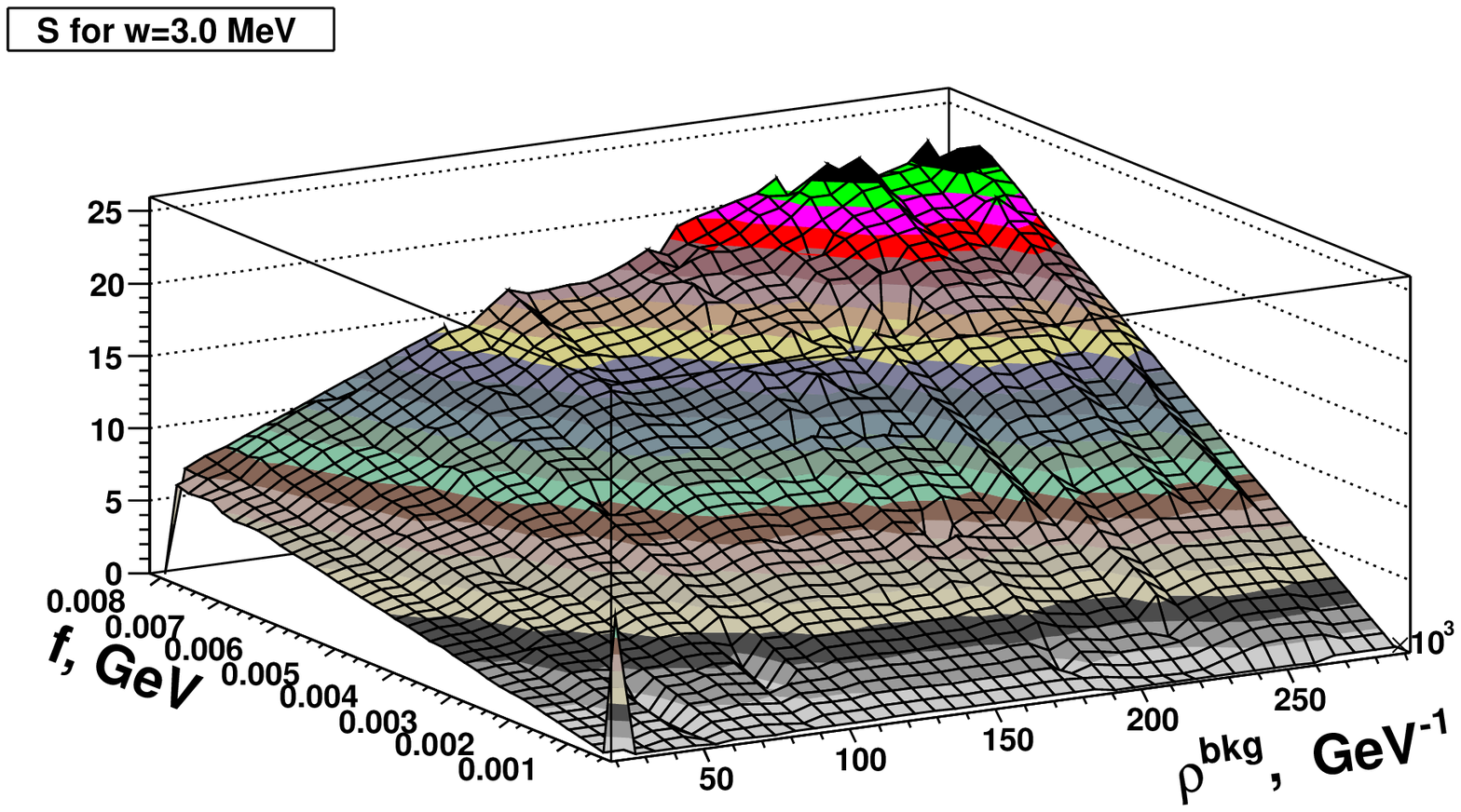,height=4.5cm}}
\end{minipage}
\hfill
\begin{minipage}[c]{.48\textwidth}
\mbox{\epsfig{file=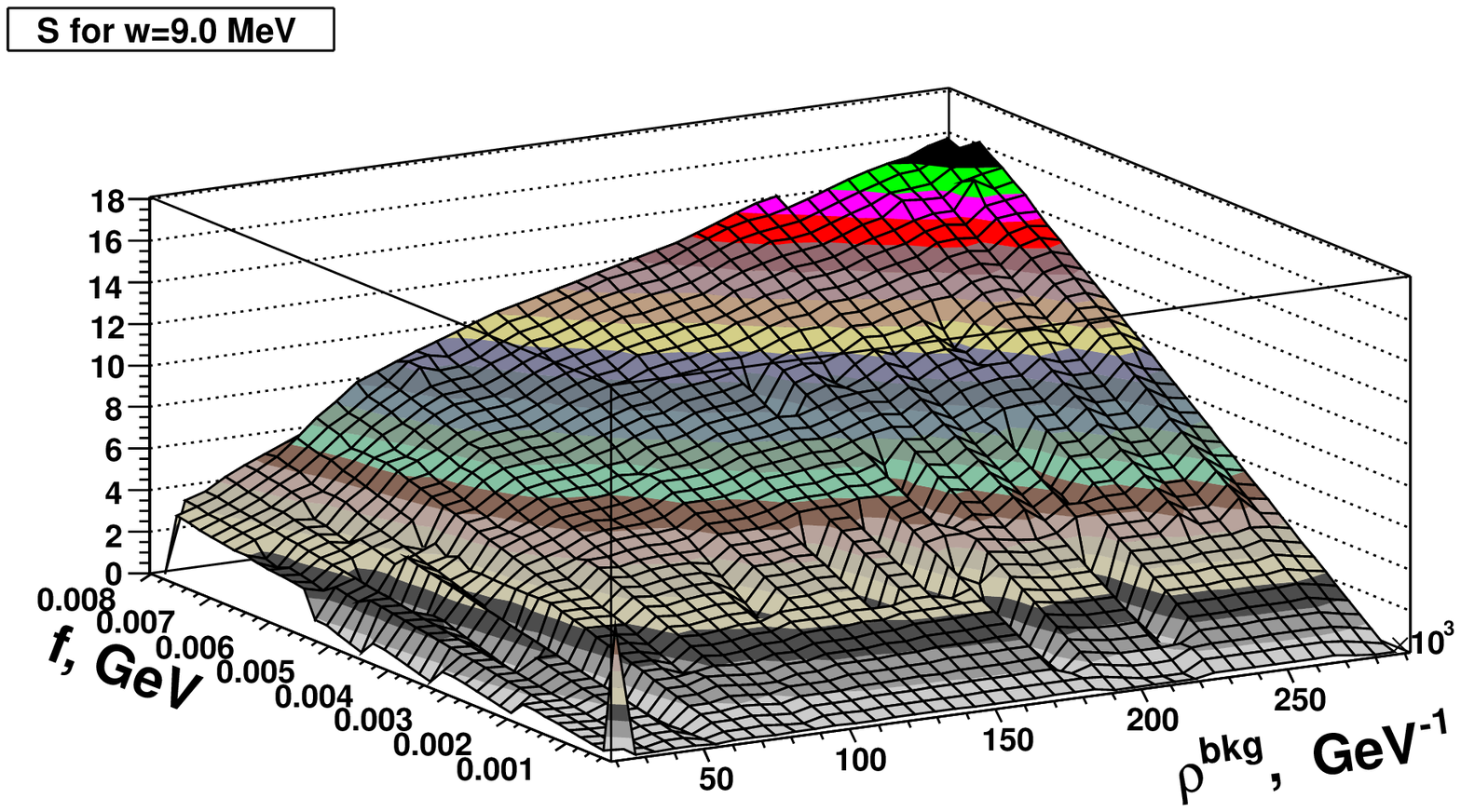,height=4.5cm}}
\end{minipage}
\begin{minipage}[c]{0.48\textwidth}
\mbox{\epsfig{file=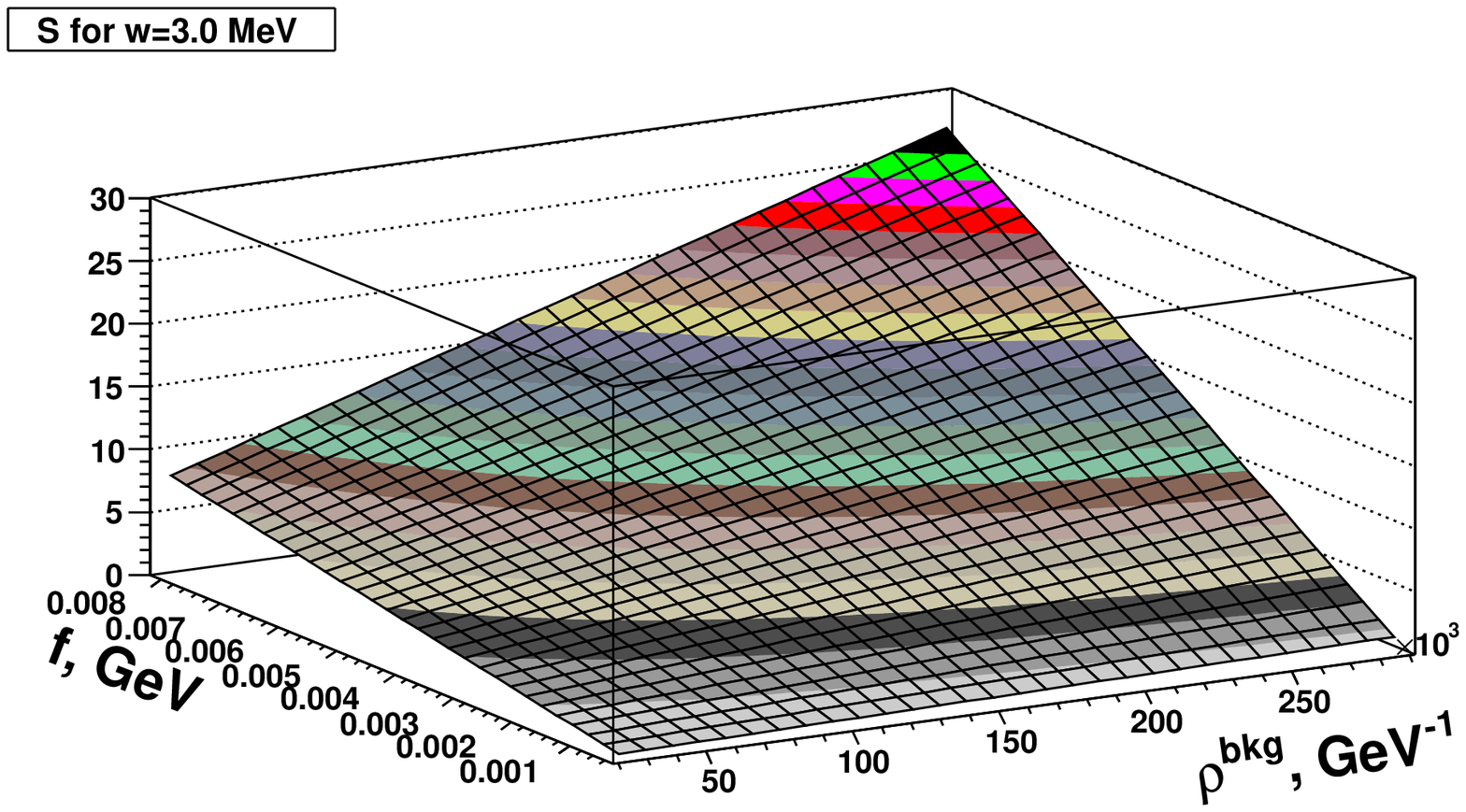,height=4.5cm}}
\end{minipage}
\hfill
\begin{minipage}[c]{.48\textwidth}
\mbox{\epsfig{file=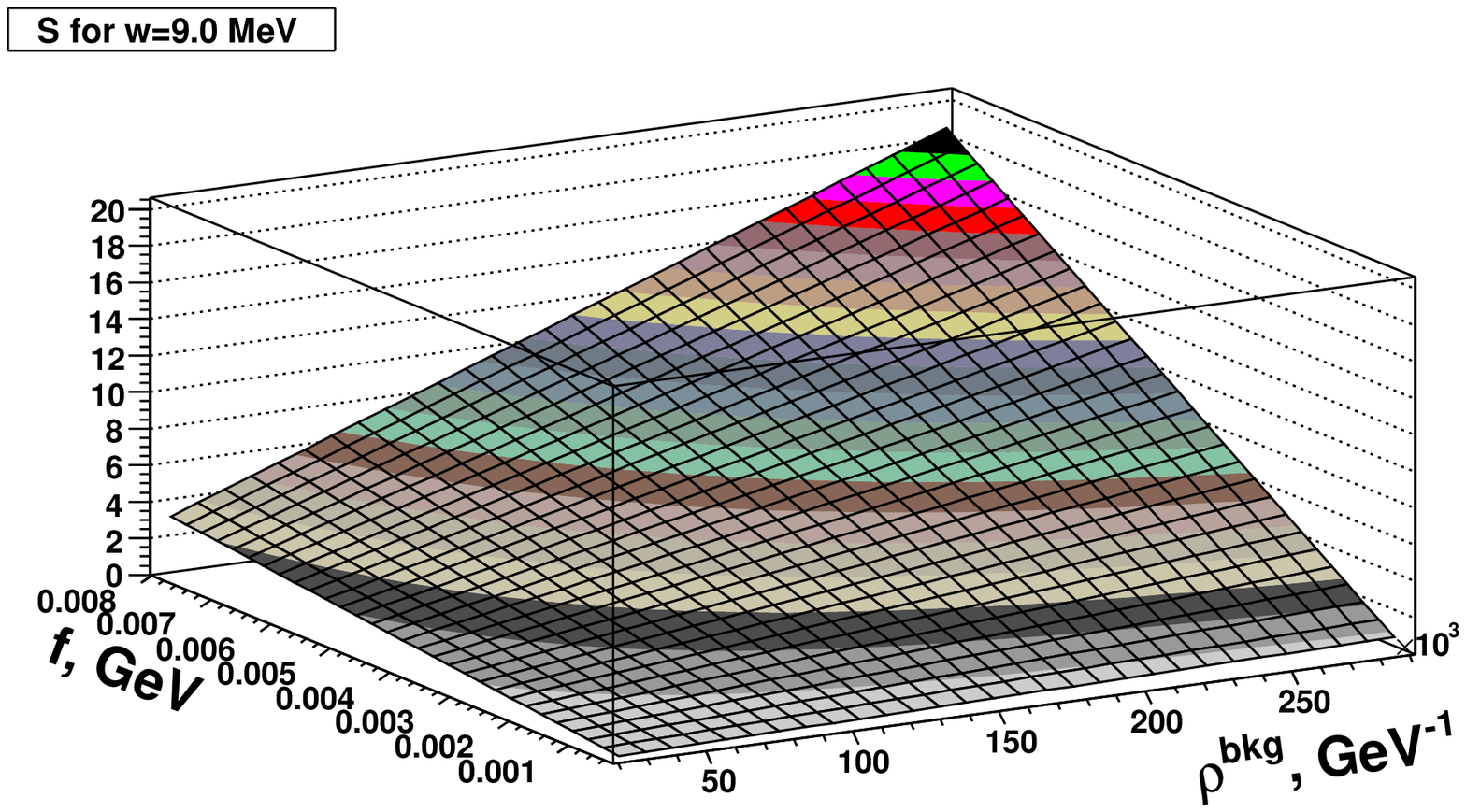,height=4.5cm}}
\label{theta1}
\end{minipage}
\caption
{The reconstructed statistical significance $S$ of a Gaussian signal as a function of the
signal fraction $f$ and the background density $\bg$ under the signal peak for
the signal width of  $3$ and $9\mev$, respectively.  The observed wrinkles are due to unstable
fits ($S$ was set to zero in such cases).
Bottom: 2D fits of the distributions shown above using the
fit function $p_1 f  + p_2 f \bg$.
Bins with zero numbers of entries were not included.
}
\label{fig2}
\end{center}
\end{figure}

Figure~\ref{fig2} shows the calculated statistical significance as a
function of $f$ and $\bg$ for 3 MeV and 9 MeV widths,
respectively.
The statistical significance was estimated as for Fig.~\ref{fig1}, i.e.
using a fit with the Gaussian distribution to extract the number of the 
signal events.
As expected, $S$ increases with increase of $f$ and $\bg$.
The irregularities seen in Fig.~\ref{fig2} are due to
unstable fits, when the resulting Gaussian width
is different from the expected width by a factor two.
In such cases, the statistical significance was set to zero.
The fraction of unstable fits was $\simeq 6\%$, and  it increases
at small $\bg$ and $f$.

For a fixed peak width $w$, the statistical significance 
as a function of the variables $f$ and $\bg$
can be fitted using the function:

\begin{equation}
S  = p_1 f  + p_2 f \bg.
\label{eq}
\end{equation}
Figure~\ref{fig2} shows the fit results using the above function (bottom plots).
The function gives a reasonable fit with typical values 
for the $\chi^2/\mathrm{ndf}$  around 0.9--1.4.

Next, $p_1$ and $p_2$ as functions of $w$  were fitted using a second
order polynomial. This ultimately leads  to the following parameterisation
of the statistical significance as functions of $f$, $\bg$ and $w$ 

\begin{equation}
S  = c_0 + c_1 w + c_2 w^2, \qquad c_i=f (a_i + b_i \bg)
\label{eq1e}
\end{equation}
where $a_0 =  1440$, $a_1 = -61000$, $a_2=0$, $b_0 =  0.0115$, 
$b_1 = -1.1$ and $b_2=44$ (here we drop unites for simplicity). 
This parameterisation is expected to be correct within $20\%$ accuracy in a region 
close to the usual discovery threshold of $5\sigma$, and when $w$ is much smaller
than the convex radius of the background shape.   
For very large values of $S$ or a very complex background shape it may fail.
Using Eq.~(\ref{eq1e}), one can easily reproduce the significance numbers
shown in Figure~\ref{fig1}.  
 
Thus, for a known background density, the expected signal fraction (or cross section)
and the expected signal width (which could be  due to the detector resolution,
a natural peak width or combinations of both), one can predict the expected signal
significance for a convex-like  background.
This does not require a Monte Carlo simulation.

\section{Analysis of similar experiments searching \\ for the $p\ks$ bump}
\label{pq}

The previous analysis can  directly be applied
to the experiments searching for the $\tp$  state.
Let us consider the case when the production cross
sections are expected to be the same, while two experiments have different
conclusions on the existence of the $\tp$  state, as in case of the 
ZEUS and H1 experiments.
The ZEUS experiment~\cite{Chekanov:2004kn} observes a narrow peak near 1522 GeV,
while H1 does not see it~\cite{Aktas:2006ic}.
In ZEUS case, $\bg=270/0.005~\gev^{-1}$ and $f=155/54000~\gev$,
assuming the most conservative single-Gaussian fit which does
not take into account the background shape near the peak\footnote{The double-Gaussian fit leads to a larger number of candidates and to a higher statistical significance.}.
The observed  $\tp$  width  was  $\sim 5$ MeV  (see Table 1 in~\cite{Chekanov:2004kn}).

At this moment, it is only possible to compare ZEUS and H1 resolutions from Monte Carlo simulations.  
For ZEUS, the $p\ks$ resolution is $2\mev$~\cite{Chekanov:2004kn}, 
while it is close to $6\mev$~\cite{private,Aktas:2006ic} near $1.52\mev$ for H1.  
For a conservative estimate, we assume that the Monte Carlo simulation
turns to underestimate the resolution and the $5\mev$ width observed by ZEUS is totally accounted
for by the tracking resolution (the natural width of $\tp$ is tiny in this case).
Using the above input values and the parameterisation  Eq.~(\ref{eq1e}), one
obtains  the statistical significance of $4.2~\sigma$.
This is close to the statistical
significance for the $\tp$  quoted in the ZEUS paper,
taken into account the experimental uncertainty on the extracted width.

In H1 case, $\ks$ width is  a factor two larger than in ZEUS case,
and this can lead to the $p\ks$ mass resolution close to $10\mev$ (see Table~1).  
Again, assuming that $\tp$ has a tiny width, this resolution number should define
the peak width.  
If one sets the background
density as in H1 case, $\bg=220/0.005 \gev^{-1}$~\cite{Aktas:2006ic}, 
while keeping $f$ as 
in the ZEUS case,
one obtains $\simeq 2.9~\sigma$, which is low statistical significance observation.
It has to be also noted that the fraction of unstable fits is 
around $10\%$  for the width of 10 MeV,
while this fraction is $6\%$ in case of 5 MeV width.
This means that the fit will not converge at all in $10\%$ cases, 
even when the peak position is fixed to the expected value during the Gaussian fit.

When H1 uses the ZEUS cut on the proton momentum $p<1.5\mev$ 
to increase the proton purity~\cite{Aktas:2006ic},
this leads to a significant decrease in the available statistics for H1: In this case,
$\bg=70/0.005~\gev^{-1}$. With the same $f$ as in ZEUS case,
one obtains the statistical significance of $2.5~\sigma$ in accordance 
with Eq.~(\ref{eq1e}).
This value is similar to that estimated for the larger $\bg$.
The small sensitivity of the significance
to the background density is due to a smallness of the second term in Eq.~(\ref{eq}) when $\bg<60000\gev^{-1}$. 

Thus, the $\tp$ signal can easily be missed in H1 case, even in the simplest case when
the expected cross section, detector acceptance and statistics are
the same as for ZEUS. Indeed, 
the limit on the $\tp$ cross section which has been set by H1~\cite{Aktas:2006ic} 
is similar to the $\tp$ cross section measured by ZEUS~\cite{tpcross}.

The situation could be different if the reconstructed $\tp$  width is 
mainly determined by the natural width
of $\tp$. In this case, the differences between the H1 and ZEUS tracking are not so
important.

A similar consideration is applicable for the SVD-2 and SPHINX experiments
\cite{Aleev:2004sa,Antipov:2004jz}.
The SVD-2 experiment has the mass resolutions almost factor of two better than SPHINX,
see Table~\ref{tab2}. Therefore, if all other experimental conditions
are similar, the statistical significance for the $\tp$ state is
expected to be higher for the SVD-2  experiment.

\section{Conclusions}

We have shown how the width of a Gaussian-like signal is related to the 
observed statistical significance 
in case of a large convex-like background. 
This observation has a direct consequence for many
experiments if the reconstructed width of the signal is 
mainly determined by
the tracking resolution. With the exception of 
$e^+e^-$ colliding experiments in which 
the baryon production is expected to be 
more suppressed compared to proton-initiated reactions, 
at present, there are two groups of similar experiments which have opposite conclusions 
on the existence of $\tp$ state in the $p\ks$ channel:
H1 and ZEUS (at DESY)~\cite{Chekanov:2004kn,Aktas:2006ic}
and SVD-2 and SPHINX (at IHEP)~\cite{Aleev:2004sa,Antipov:2004jz}. 
We have shown that the ZEUS and SVD-2 experiments, which
observe a peak near 1530 MeV, have better mass resolutions 
for $\ks$($\Lambda$) and
$p\ks$ invariant masses than the H1 and SPHINX experiments.
Taking into account the observed relationship between the signal width
and the tracking resolution, this fact increases the chances of ZEUS and SVD-2  
to find the narrow signal.
We have shown that the worsening of the tracking 
resolution for the $\tp$ searches 
is the penalty the experiments have to pay if 
a significant inactive material is located in front of tracking devices.

It has to be noted that this  approach cannot be applied blindly
to all experiments which observe or do not observe the $\tp$ signal, since
such experiments may have very different acceptance,
background shape and the  production mechanism
in the reaction under study.
The approach could help to explain differences
in experimental results if the kinematics and the 
production cross section are known
to be similar.

\section*{Acknowledgements}
\vspace{0.3cm}
The authors are grateful to P.F.~Ermolov,  E.~Lohrmann and K.~Daum for reading the paper draft,
comments and useful discussions. The work of B.B.L. is partially supported
by the Russian Foundation for Basic Research under grant no. 
05-02-39028gfen-a.

\bibliographystyle{./l4z_pl}
\def\bibname{\Large\bf References}
\def\refname{\Large\bf References}
\pagestyle{plain}
\bibliography{bibfile}

\newpage

\begin{table}
\centering
\begin{tabular}{|l|c|c|c|c|}
\hline
 $L$~(cm) &   $w(K_s),$ MeV &  $w(\Theta ^+_{r}),$ MeV &
 $w(\Theta ^+_{n} ),$ MeV&   $w(\Theta ^+_{tr}),$  MeV \\
\multicolumn{1}{|c|}{}&&&&\\  \cline{1-2}
\hline
 0.3   & 4.23  & 3.84 &2.93 &  1.38 \\
 0.4   &  5.10 & 4.89 &3.38 &  1.57 \\
 0.5   &  5.79 & 5.79 &3.99 &  1.82 \\
 0.6   &  6.31 & 6.78 &4.32 &  1.84 \\
 0.7   &  6.60 & 7.43 & 5.49&  1.90 \\
 0.8   &  7.69 & 8.82 &5.19 &  2.05 \\
 0.9   &  8.05 & 9.65 &5.95 &  2.09\\
 1.0   &  8.33 & 11.46&5.97 &  2.28\\
\hline
\end{tabular}
\caption{
The reconstructed width 
(see definition in the 
text)
of $\ks$ and  $\tp$
as a function of the thickness, $L$, of an aluminum
target (in cm). The reconstruction of  $\tp$
was performed by three different methods:
1) $\ks$ is reconstructed by using tracks 
(second column)  and combined with the proton
track ($w(\Theta ^+_{r}),$ third column).
2) Pion tracks from a $\ks$ decay are combined with the proton
track, while the $p\ks$ was calculated  assuming  
the nominal $\ks$ mass ($w(\Theta ^+_{n}),$  fourth column).
3)  True charged pions from $\ks$ and the 
proton track are combined  ($w(\Theta ^+_{tr}),$  last column).
The reconstruction was performed using a {\sc Geant}  simulation
with the initial momenta of $\tp$ of 1 GeV.
}
\label{tab1}
\end{table}

\begin{table}
\centering
\begin{tabular}{|l|c|l|c|}
\hline
Experiment &   reaction &  width, MeV & $\tp$ signal \\
\multicolumn{1}{|c|}{}&&&\\  \cline{1-2}
\hline
ZEUS\cite{Chekanov:2003wc}   &   & $w(\ks )=$ 4.0         &  Yes \\
           & $ep\rightarrow K_s p +X$, $ \sqrt{s}$=300 GeV          & $w(\Lambda )\ \ =$ 1.9 &  \\
H1\cite{Adloff:1997ym}       &   & $w(\ks )=$ 9.2        &  No \\
           &           & $w(\Lambda )\ \ =$ 2.9 &   \\
\hline
SVD-2\cite{Aleev:2004sa}   &   & $w(\ks )=4.40\pm$ 0.08 &   Yes\\
          & $pA \rightarrow K_s p +X,$ 70 GeV& $w(\Lambda )\ \ =1.60\pm$ 0.04  &   \\
SPHINX\cite{Antipov:2004jz}  &   & $w(\ks )=$ 8.4  &  No \\
   &   & $w(\Lambda )\ \ =$ 3.8   &   \\
\hline
\end{tabular}
\caption{A compilation of $\ks$ and $\Lambda$ mass resolutions, $w$,
in the two groups of  experiments with opposite conclusions on the  $\tp$ existence.
The width $w$ denotes the standard 
deviation of a Gaussian distribution (see the text).
}
\label{tab2}
\end{table}

\end{document}